\documentclass[twoside,11pt]{article}

\usepackage[abbrvbib]{jmlr2e}

\jmlrheading{}{2020}{}{}{}{}{Ankit Sharma, Garima Gupta, Ranjitha Prasad, Arnab Chatterjee, Lovekesh Vig and Gautam Shroff.}

\ShortHeadings{Hi-CI: Deep Causal Inference in High Dimensions}{Sharma, Gupta, Prasad, Chatterjee, Vig and Shroff.}
\firstpageno{1}

\begin{document}

\title{Hi-CI: Deep Causal Inference in High Dimensions}

\author{\name Ankit Sharma \email ankit.sharma16@tcs.com \\
       \addr TCS Research, India
       \AND
       \name Garima Gupta \email gupta.garima1@tcs.com \\
       \addr TCS Research, India
       \AND
       \name Ranjitha Prasad \email ranjitha@iiitd.ac.in \\
       \addr Indraprastha Institute of Information Technology, Delhi, India
       \AND Arnab Chatterjee \email arnab.chatterjee4@tcs.com \\
       \addr TCS Research, India 
       \AND Lovekesh Vig \email
       lovekesh.vig@tcs.com \\ 
       \addr TCS Research, India 
       \AND Gautam Shroff \email
       gautam.shroff@tcs.com \\
       \addr TCS Research, India}

\editor{}

\maketitle

\begin{abstract}
 We address the problem of counterfactual regression using causal inference (CI) in observational studies consisting of high dimensional covariates and high cardinality treatments. Confounding bias, which leads  to inaccurate treatment effect estimation, is attributed to covariates that affect both treatments and outcome. The presence of high-dimensional covariates exacerbates the impact of bias as it is harder to isolate and measure the impact of these \emph{confounders}. In the presence of high-cardinality treatment variables, CI is rendered ill-posed due to the increase in the number of counterfactual outcomes to be predicted. We propose \texttt{Hi-CI}, a deep neural network (DNN) based framework for estimating causal effects in the presence of large number of covariates, and high-cardinal and continuous treatment variables. The proposed architecture comprises of a decorrelation network and an outcome prediction network. In the decorrelation network, we learn a data representation in lower dimensions as compared to the original covariates, and addresses confounding bias alongside. Subsequently, in the outcome prediction network, we learn an \emph{embedding} of high-cardinality and continuous treatments, jointly with the data representation. We demonstrate the efficacy of causal effect prediction of the proposed \texttt{Hi-CI} network using synthetic and real-world NEWS datasets.
\end{abstract}
\begin{keywords}
  Causal Inference, Confounding, High-dimensions, Deep neural networks, Embedding
\end{keywords}

\section{Introduction}

In our daily lives, humans often justify several actions and events in terms of \emph{cause and effect} and incorporate the feedback into future actions in order to alter future outcomes. On the other hand, supervised learning based machine learning techniques focus on predicting outcomes, where outcomes are strongly tied to the nature of training data. However, it is possible that when this model is used in real-life scenarios, data generating process may vary vastly, and hence these models do not generalize well. Hence, there have been several efforts by researchers to integrate causality into machine learning models for obtaining robust and generalizable machine learning models. It is well-accepted that obtaining causal relations from an observational dataset is possible if underlying data generating process is well-understood. This is often posed as a problem of predicting the effects of interventions (or \emph{treatments}) in the data generating process, and such treatments are generally enforced using policy or operational changes. Furthermore, understanding the effect of intervention requires us accurately answer counterfactual or \emph{what-if} type questions, which in turn necessitates modelling the  causal relationship between the treatment and outcome variables.

Causal inference (CI) for observational studies lies at the heart of various domains like healthcare, digital marketing, econometrics based applications, etc, that require quantifying the effect of a treatment or an intervention on an individual. As an example, consider a retail outlet optimizing the waiting time at a store since long queues leads to loss in customer base, in turn leading to low sales. In their historical observational data, consider the queue-length as a treatment variable and sale as an outcome variable. First, note that queue-length varies in the training data since it depends on the number of items purchased by every customer. A discount sale leads to a given customer buying more leading to higher queue-length. That is, in our training set we observe examples with long queues and high sales. A naive supervised learning approach might incorrectly predict that increase in queue-length leads to increase in sales, whereas the true relationship between queue-length and sales is surely negative on regular days. Typically we have information regarding discount sales, and including them in the model can correct for such effects. Such variables affect both, the outcome and the treatment, and hence, these variables are known as \emph{confounding covariates} in the CI problem. Similarly, in a digital marketing context, \emph{age} can be a confounding covariate which introduces selection bias in providing advertisements to young, middle-aged and old-aged users and consequently a varying buying behaviour (outcome). These aspects as well-captured in Simpson's paradox~\citep{bottou2013counterfactual}, which states that the confounding behavior may lead to erroneous conclusions about causal relations and counterfactual estimation when the confounding variable is not considered in analysis. 

 A key problem in modern empirical work is that datasets consists of large numbers of covariates~\citep{Newman-2012} and high-cardinality  treatments~\citep{DiemertMeynet2017}. We motivate this using the following real-world scenarios
 \begin{enumerate}
     \item High-dimensional covariates: A typical characteristic of genomic data is the presence of vast number of covariates. For example, a problem of interest is to genetically modify the plant \emph{Arabidopsis thaliana} to shorten the time to flowering \citep{buhlmann2013causal} since fast growing crops lead to better food production. In the corresponding dataset, there are $47$ instances of the outcome \emph{time to flowering} and $21,326$ genes which are construed as \emph{covariates}. The goal is to causally infer the effects of a single gene intervention on the outcome, considering the other genes as the covaraites. A similar (but less severe) situation is also seen in the popular \emph{The Cancer Genomic Atlas} (TCGA) project \citep{weinstein2013cancer} which is a repository that consists of gene expression values of $20547$ genes of $9659$ individuals. Here the goal is to measure the gene expression values for several treatment strategies like medication, chemotherapy and surgery \citep{schwab2019learning}, so that the best treatment regimen is chosen. 
     \item High-cardinality treatments: We provide the example of the Criteo dataset to motivate high cardinality treatments. Criteo dataset~\citep{DiemertMeynet2017} includes browsing related activities of users for interaction with $675$ campaigns. In the causal setting, these campaigns are considered as \emph{treatments} with \emph{campaign effect on buying} as the outcome~\citep{dalessandro2012causally}.
     \item High dimensional covariates, high cardinality treatments with dosages: The popular NEWS datasets consists of news items represented by $2870$ bag-of-word covariates. These news items are read by viewers on media devices. In causal setting, media devices act as treatments. Since the number of news items can vary from few tens to hundreds, varying but finite viewing time is considered as dosage levels, while the readers' opinion on different media devices is considered as outcome \citep{schwab2019learning}.
 \end{enumerate}
 
In the above applications of healthcare, advertising etc, an individual's response plays an important role in guiding practitioners to select the best possible interventions. Hence, it is essential to build models to handle such high dimensional scenarios. This motivates us to design machine learning models that abate confounding effects, while being parsimonious in representation of high-dimensional variables, and adequately flexible. 

\begin{figure*}[t]
\centering
\includegraphics[width=0.82\textwidth]{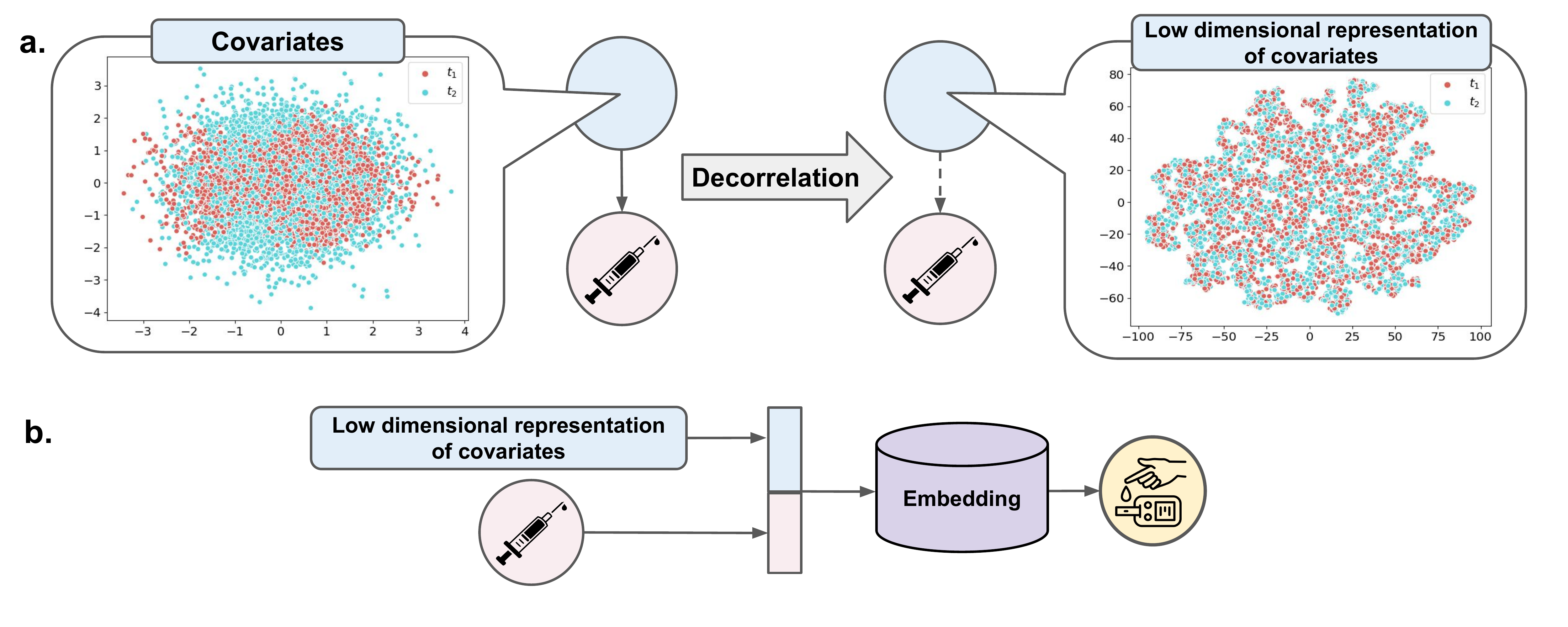}
\caption{\textbf{a.} The t-SNE plot on the left and right depicts the decorrelated transformation of high-dimensional covariates in data, into a low-dimensional representation, using the \texttt{Hi-CI} framework.  \textbf{b.} Illustrates the dosage embedding to learn a low-dimensional representation of treatments followed by outcome prediction in the \texttt{Hi-CI} framework. }
\label{fig:introFig}
\end{figure*}

\textbf{Related Work:} ~~Treatment effect estimation in the presence of high dimensional covariates is a well-explored topic in statistical literature on causal inference. In \citep{robins1994estimation}, the authors propose techniques based on inverse probability of treatment weighting (IPTW), which is sensitive to the propensity score model~\citep{fan2016improving}. Propensity score estimation is improved by employing covariate balancing propensity scores (CBPS) in high dimensions ~\citep{imai2014covariate,guo2016cbps,fan2016improving}. LASSO regression for high dimensional CI was proposed in \citep{belloni2014inference}. Approximate residual balancing techniques for treatment effect estimation in high dimensions is proposed in \citep{athey2018approximate}. A common trait among these works is that they focus on estimating the average treatment effect (ATE) in the presence of a large number of covariates, but are limited to settings with only two treatments. In \citep{schwab2019learning}, high-cardinality treatments and continuous treatments are considered. Typically, in the context of continuous treatments, a given treatment is represented using \emph{multiple dosage levels} \citep{schwab2019learning} to account for the exploding cardinality of the treatment set (as each dosage is a unique treatment in itself). In statistical literature, continuous dosages have been handled using propensity scores~\citep{hirano2004propensity}, doubly robust estimation methods~\citep{kennedy2017non}, generalized CBPS score~\citep{fong2018covariate}, using estimation frameworks for both treatment assignment and outcome prediction~\citep{galagate2016causal}.

Modern deep neural networks (DNN) based methods employ matching or balancing techniques for compensating confounding bias. Existing DNN based architectures for the multiple treatment scenario as proposed in ~\citep{esann-multiMBNN,schwab2018perfect} have a severe limitation with respect to their architectures. They employ a  separate regression network per treatment, and hence, these neural networks cannot be used in the presence of a large number of treatments. Furthermore, in the presence of high-dimensional covariates, it is essential to design a parsimonious, yet lossless representation of these covariates. In several works such as~\citep{johansson2016learning,pmlr-v70-shalit17a}, a latent representation for covariates is learned by minimizing the discrepancy distances of the control and treatment populations to compensate for confounding bias, in the presence of binary treatments. Since such a data representation is not lossless, this approach is not suitable in the presence of high-cardinality variables. An autoencoder is used to learn an unbiased lossless representation of covariates, uncorrelated with respect to the multiple, yet small number of treatment variables ~\citep{atan2018deep,zhang2019reducing}. On the other hand, matching based DNN techniques find similar individuals with dissimilar treatments using propensity scores~\citep{schwab2018perfect,esann-multiMBNN,ho2007matching}. Matching is often accomplished using nearest neighbour match~\citep{ho2007matching}, propensity score~\citep{schwab2018perfect} or generalized propensity score~\citep{esann-multiMBNN}. These techniques are computationally infeasible in the presence of high-cardinality treatment variables as good recipes for matching require spanning the entire dataset in search of alternate treatment variables while ensuring a balance in the number of individuals per treatment. However, the success of the DNN based causal CI techniques and vacuum in the literature on architectures that handle large number of covariates and treatments in observational datasets motivates us to address this problem.

 \textbf{Contributions}: In the presence of high-cardinality treatment variables, the number of counterfactual outcomes to be estimated is much larger than the number of factual observations, rendering the problem to be ill-posed. Furthermore, lack of information regarding the confounders among large number of covariates pose challenges in handling confounding bias. Hence, it becomes essential to find a lower dimensional manifold where an equivalent problem of causal inference can be posed, and counterfactual outcomes can be computed. To this end, we propose a DNN based \texttt{Hi-CI}, specifically designed to measure causal effects in the presence of countably large number of covariates and high-cardinality treatments. We propose a novel loss function which is designed to achieve the following goals:
 \begin{itemize}
     \item Obtain an autoencoder based data representation for high-dimensional covariates while simultaneously handling confounding bias using a decorrelation loss, as depicted in (a) of Fig.~\ref{fig:introFig}.
     \item Caters to both, a large number of discrete, and  continuous treatments, where a continuous treatment is characterised by a fixed number of dosage levels.
     \item Obtain a per-dosage level embedding layer to learn the low-dimensional representation of the high-cardinality treatments by jointly training the network using root mean square (RMSE) loss and a sparsifying mixed norm loss function as depicted in lower part (b) of Fig.~\ref{fig:introFig}.
 \end{itemize}

 We demonstrate that the proposed \texttt{Hi-CI} architecture outperforms the state-of-the-art CI techniques such as perfect match (\texttt{PM})~\citep{schwab2018perfect}, \texttt{MultiMBNN}~\citep{esann-multiMBNN}, \texttt{Deep-Treat}~\citep{atan2018deep} and \texttt{DRNet}~\citep{schwab2019learning} in the presence of high-dimensional variables. We use precision in estimation of heterogeneous effect ($PEHE$) and mean absolute percentage error ($MAPE$) over the ATE as metrics in all the experiments. For treatments with dosages, we use mean integrated square error ($MISE$) as a metric, and in addition, we also introduce a new metric for computing $MAPE$ over the dosage effect.

\section{CI in High dimensions: Preliminaries and Loss Function}
In this section, we describe the causal inference problem, especially in the presence of a large number of covariates and treatments. We motivate the need for finding a low-dimensional representation of the high-cardinality covariates and treatments and describe the loss function that helps to learn these representations from data and hence provide the basis for understanding the \texttt{Hi-CI} DNN described in the sequel.
\subsection{Causal Inference Preliminaries}
\label{sec:prelim}
We consider training data $\matD_{CI}$, comprising of $N$ samples from an observational dataset, where each sample is given by $\{x_n,\vect_n,\vecy_{n}\}$, where $x_n \in \mathbf{X}$. Each individual (also called \textit{context}) $n$ is represented using $P$ covariates, i.e., $x_{n,p}$ denotes the $p$-th covariate of the $n$-th individual, for $1\leq n \leq N$. Furthermore, an individual is subject to one of the $K$ treatments given by $\vect_n = [t_n(1), \hdots, t_n(K)]$, where each entry of $\vect_n$ is binary, i.e., $t_n(k) \in \{0,1\}$. Here, $t_n(k) = 1$ implies that the $k$-th treatment is provided. We assume that only one treatment is provided to an individual at any given point in time, and hence, $\vect_n$ is an one-hot vector. A \emph{counterfactual} is defined based on $K-1$ alternate treatments, and corresponding outcomes are referred to as counterfactual outcomes. Accordingly, the response vector for the $n$-th individual is given by $\vecy_{n} \in \mathbb{R}^{K \times 1}$, i.e., the outcome is a continuous random vector with $K$ entries denoted by $y_{n}(k)$, the response of the $n$-th individual to the $k$-th treatment. The set of counterfactual responses for the $n$-th individual comprises of response to treatments $l \neq k$, given by $y_{n,l}$, and the size of this set is $K-1$. In the case of continuous treatment, we assume that $t_n(k) \in \mathbb{R}$ which implies that $\vect_n \in \mathbb{R}^{K \times 1}$, that is the treatment is a real-valued vector. However, for ease of representation, we cast the continuous treatment variable using a finite set of $E$ dosage levels where $E$ remains constant across treatments. Following the notation for discrete treatments, the outcome is a continuous random vector denoted by $y_{n}(k_e)$, where $1 \leq k_e \leq KE$, is the response of the $n$-th individual to the $e$-th dosage level of the $k$-th treatment. In the case of discrete treatment, the maximum size of outcomes to be predicted by the DNN is $N(K-1)$, while the number of available factual outcomes are $N$ in number. It is evident that this problem is ill-posed when $K$ is large. Furthermore, in the case of continuous treatments, we effectively have $KE$ treatments, leading to $N(KE-1)$ counterfactual responses. 

In this work, we consider observational studies where there are large number of covariates $P$ and large number of treatments $K$. Our goal is to train a DNN model to overcome confounding and perform counterfactual regression, i.e., to predict the response given any context and treatment, for large $P$ and $K$. In the sequel, we describe different components of the the loss function that tackles confounding bias, high-dimensional treatments and high-dimensional covariates.  

\subsection{Learning Representations from Data}
\label{sec:balancing}
The crux of the loss function in CI for observational studies lie in techniques employed to compensate for the confounding bias. In this direction, we employ autoencoders which simultaneously encourage confounding bias compensation and learning compressed representation for the high-dimensional covariates. Alongside, we employ RMSE with mixed-norm regularizer based loss-function to obtain a low-dimensional representation for treatments. In the sequel, we describe the mathematical constructs of learning the representation and the loss function. 

\subsubsection{Data Representation using autoencoders}
In this work, we propose to use the autoencoder to jointly obtain a low-dimensional representation of the high-dimensional covariates and alleviate the effect of confounding. Let $\matT$ represent the set of treatments, and $T_k \in \matT$ be a random variable; instantiation for the $n$-th individual is $t_n(k)$. Using an autoencoder, we seek a mapping from the space of covariates $\matX \in \mathbb{R}^{N \times P}$, such that $\Phi: \matX \rightarrow \mathcal{R}$, where $\mathcal{R} \in \mathbb{R}^{N \times L}$ is the representation space. The mapping $\Phi$ is such that,
\begin{itemize}
    \item The induced distribution of the treatments over $\mathcal{R}$, which we denote $p(T_k |\Phi(\matX))$ is free of confounding bias for all $k$. 
    \item The representation of $\vecx_n$ under $\Phi(\cdot)$ for all $n$ is lossless.
    \item It maps higher dimensional covariates in $P$ to a low-dimensional space of size $L$, i.e., $L < P$.
\end{itemize}

A typical propensity score based matching approach addresses the issue of confounding bias by balancing the propensity score to obtain similar covariate distributions across treated populations. Mathematically, a sub-sample $\matX_s$ of the original sample is considered such that it ensures that the following condition holds:
\begin{equation}
    p(T_1|\matX_s) = p(T_2|\matX_s) = \hdots = p(T_K|\matX_s).
\end{equation}
Note that the condition stated above does not necessitate that treatment and covariates variables are uncorrelated. On the other hand, the loss function associated to the autoencoder imposes a far more stringent condition~\citep{atan2018deep} such that
\begin{equation}
    p(T_k|\matX) = p(T_k),~~ \forall k,
\end{equation}
for the entire sample $\matD_{CI}$. Autoencoders have been employed in the literature for addressing some of the tasks such as lossless data representation~\citep{atan2018deep,ramachandra2018deep}. However, to the best of the authors' knowledge, this is the first-of-its-kind approach where an autoencoder is used to jointly accomplish the goals as specified above, and primarily, low-dimensional representations. 

To ensure lossless data representation, the loss function associated with the autoencoder jointly minimizes the mean-squared error loss between the reconstructed and the original covariates, and the distance between the unbiased ($p(T_k)$) and the biased treatment distributions ($p(T_k|\Phi(\mathbf{X}))$) for all $k$, while maintaining the resultant mapping in a lower dimension as compared to the original covariates ($L < P$). These goals can be achieved by using the following loss function:
\begin{equation}
    \mathcal{L}_1(\Phi,\Psi,\beta) = \mathcal{L}_{ce}(\Phi(\matX)) + \beta \mathcal{L}_{ae}(\Phi(\matX),\Psi(\Phi(\matX))),
\end{equation}
where $\mathcal{L}_{ce}(\Phi)$ is the cross-entropy measure. The cross-entropy loss is directly proportional to the Kullback-Liebler divergence between the distributions in question, and hence it is an appropriate metric to minimize the divergence between $p(T_k)$ and $p(T_k|\Phi(\matX))$ for all $k$. Accordingly, $\mathcal{L}_{ce}(\Phi)$ is given by
\begin{equation}
    \mathcal{L}_{ce}(\Phi)  = -\sum_{T \in \mathbf{T}} p(T)\log(p(T|\Phi(\matX))).
\end{equation}
Furthermore, the loss term $\mathcal{L}_{ae}(\Phi,\Psi)$ is employed to minimize the mean-squared loss between the reconstructed and the original covariates in the autoencoder. Mathematically, 
\begin{equation}
    \mathcal{L}_{ae}(\Phi,\Psi) = \frac{1}{PN}\sum_{n = 1}^N\sum_{p = 1}^P\lvert\lvert(\vecx_{n,p} - (\Phi\circ\Psi)(\vecx_{n,p}))\rvert\rvert^2,
\end{equation}
where $\Psi$ is the decoder mapping such that $\Psi: \mathcal{R} \rightarrow \matX$ and $\circ$ is a composition operator, and $L < P$, which ensures that we obtain a low-dimensional, yet meaningful representation of the high-dimensional covariates. 

As a regularizer, we propose to employ the mixed norm on the difference of means, represented using the matrix $\matM_D$. The columns of $\matM_D$ are given by $\mu_{D,(T_i,T_j)} = \frac{1}{LK(K-1)}(\mu_{T_i}(\Phi(\matX)) - \mu_{T_j}(\Phi(\matX)))$, where $\mu_{T_i}(\Phi(\matX)) \in \mathbb{R}^{L \times 1}$ is the mean of representation for all individuals in $\matX$, given by $\Phi(\matX)$, that undergo treatment $T_i$. Since all possible pairs of treatments $(T_i,T_j)$, for all $T_i$ and $T_j$ are considered, $\matM_D$ is of dimension $\mathbb{R}^{L \times (K(K-1))}$. The mixed norm regularizer on $\matM_{D}$, denoted as $\mathcal{L}_{2,1}(\matM_D)$, is as follows:
\begin{equation}
    \mathcal{L}_{2,1}(\matM_D) = \sum_{u = 0}^{K(K-1)}\sqrt{\sum_{v = 0}^{L-1} \lvert M_D(u,v)\rvert^2}.
    \label{eq:ell21}
\end{equation}
Combining the above mentioned loss functions, we obtain the following loss function from the decorrelating network:
\begin{equation}
    \mathcal{L}_D(\Phi,\Psi,\beta,\gamma) = \mathcal{L}_{ce}(\Phi) + \beta \mathcal{L}_{ae}(\Phi,\Psi) + \gamma \mathcal{L}_{2,1}(\matM_D).
    \label{eq:ellD}
\end{equation}
The above objective function cannot be computed directly since both $p(T_k |\Phi(\matX))$ and $p(T_k)$ are unknown for any $k$. The estimates of $p(T_k)$ for $1 < k \leq K$ is given by \citep{atan2018deep}:
\begin{equation}
    p(T_{k} = t) = \frac{\sum_{n = 1}^N \mathbb{I}(t_n(k) = t)}{N},
\end{equation}
where $\mathbb{I}(\cdot)$ is the indicator function. Essentially, $p(T_k)$ provides a count-based probability of $k$-{th} treatment. Further, the functional form of $p(T_k|\Phi(x_n))$ is assumed to be similar to logistic regression:
\begin{equation}
   p(T_k|\Phi(\vecx_n)) = \frac{\exp({(\bm{\theta}}_{T_k})^T\Phi(\vecx_n))}{\sum_{k=1}^{K}\exp({(\bm{\theta}}_{T_k})^T\Phi(\vecx_n))},
   \label{eq:logisticloss}
\end{equation}
where $\bm{\theta}_{T_{k}} \in \mathbb{R}^{L \times 1}$ are the per-treatment parameters of the logistic regression framework. This results in 
\begin{equation}
     \mathcal{L}_{ce}(\Phi) = -\sum_{k=1}^{K} p(T_k)(\bm{\theta}_{T_{k}})^T\Phi(\vecx_n) - p(T_k)\log{\left(\sum_{k=1}^{K}\exp((\bm{\theta}_{T_{k}})^T\Phi(\vecx_n))\right)}.
    \label{eq:ellce}
\end{equation}

\subsubsection{Embeddings for high-dimensional treatment}
As mentioned in the previous subsection, the goal of this work is to design DNN based CI model for datasets with large number of unique treatments. While a single bit is sufficient to represent binary treatments~\citep{johansson2016learning}, a one hot representation is used within the DNN to represent a categorical treatment for a given user~\citep{esann-multiMBNN}. In the presence of high-cardinality treatment variables, i.e., treatments with several unique categories, the size of the one-hot vector becomes unmanageable. Furthermore, DNN architectures that cater to multiple treatments often use a sub-divided network as in~\citep{schwab2018perfect} and~\citep{schwab2019learning}, with one branch per treatment. Such a branching network based DNN architecture becomes computationally intractable as the number of treatments increase. 

An aspect that matters the most about one-hot encoding is the fact that one-hot mapping does not capture any similarity in treatment categories. For instance, if treatments $t_1$ and $t_2$ are drugs for lung-related issues, and $t_3$ is a treatment for skin-acne which is seemingly an unrelated issue, $t_1$, $t_2$ and $t_3$ are equidistant in the one-hot encoding space. We propose to learn a representation of treatments denoted as $\Omega: [\Phi(\matX),\matT] \rightarrow \matY$, where $\matY$ represents output response vector, and the \emph{embedding} $\Omega$ encapsulates closeness property of treatments. Such representations of the treatment space is extremely relevant in the current day observational studies, as explained in the introduction. 

The impact of the embedding is realised in the outcome prediction part of the network. The loss on the outputs of the outcome prediction layer is the root mean square error (RMSE) loss given by
\begin{equation}
    \mathcal{L}_{RMSE}(\vecy,\hat{\vecy}) = \sqrt{\frac{1}{N}\sum_{n = 1}^N\lvert\lvert y_n - \hat{y}_n \rvert\rvert^2}
    \label{eq:lossRMSE}
\end{equation}
where $\hat{\vecy}_n = \Omega([\Phi(\vecx_n),\vect_n]^T)$.

Although the impact of embedding is evident only in the above loss function, note that the training of the \texttt{Hi-CI} framework incorporates all of the loss functions combined in \eqref{eq:ellD} and \eqref{eq:lossRMSE}. Intuitively, through the mixed norm based regularizer in \eqref{eq:ell21}, we minimize the distance between multiple populations whose covariate information is summarized by $\Phi(\matX)$ and hence, we are able to exploit the similarity properties in the treatment itself. When we train the network using \eqref{eq:lossRMSE} along with \eqref{eq:ell21}, in addition to promoting parsimonious representations owing to similarity of treatments, we also ensure that such representation leads to a response close, in the sense of RMSE, to the true label. 

\subsection{Modified Loss Function when $E > 1$}
In the case of continuous treatment, we represent a treatment as consisting of multiple dosages~\citep{schwab2019learning}. In particular, we assume that each treatment is specified by a set of $E$ dosage levels, i.e., $E$ remains constant across treatments. In the design of DNN, we assume that the treatment is affected by the confounding bias, but the dosage administered is not. However, since we need to infer the per-dosage level counterfactual, we exploit the dosage information available in the labels $\vecy_{ne}$ outcome prediction network. Accordingly, we incorporate the dosage levels in a generalised RMSE loss function: 
\begin{equation}
    \mathcal{L}_{RMSE}(\vecy,\hat{\vecy}) = \sqrt{\frac{1}{N}\sum_{n = 1}^{N}\sum_{e = 1}^E \lvert\lvert y_{ne} - \hat{y}_{ne} \rvert\rvert^2},
    \label{eq:lossRMSEDosage}
\end{equation}
where $\hat{\vecy}_n = \Omega_e([\Phi(\vecx_n),\vect_n]^T)$. Note that if we set $E = 1$ in \eqref{eq:lossRMSEDosage}, we obtain \eqref{eq:lossRMSE}. Hence, we essentially aggregate the characteristics of all observed individuals subjected to any treatment, and having been administered a given dosage $e_n$. 

In order to implement the loss function associated with continuous treatments, it is also essential to make changes in the network architecture. In the sequel, we describe the proposed DNN architecture of $\texttt{Hi-CI}$ in tandem with components of the loss-function as described in \eqref{eq:lossRMSE} and \eqref{eq:ellD}. We also discuss the changes in the network architecture due to the generalised RMSE loss in \eqref{eq:lossRMSEDosage}.

\begin{figure*}[t!]
\includegraphics[width=\linewidth, height=5.5cm]{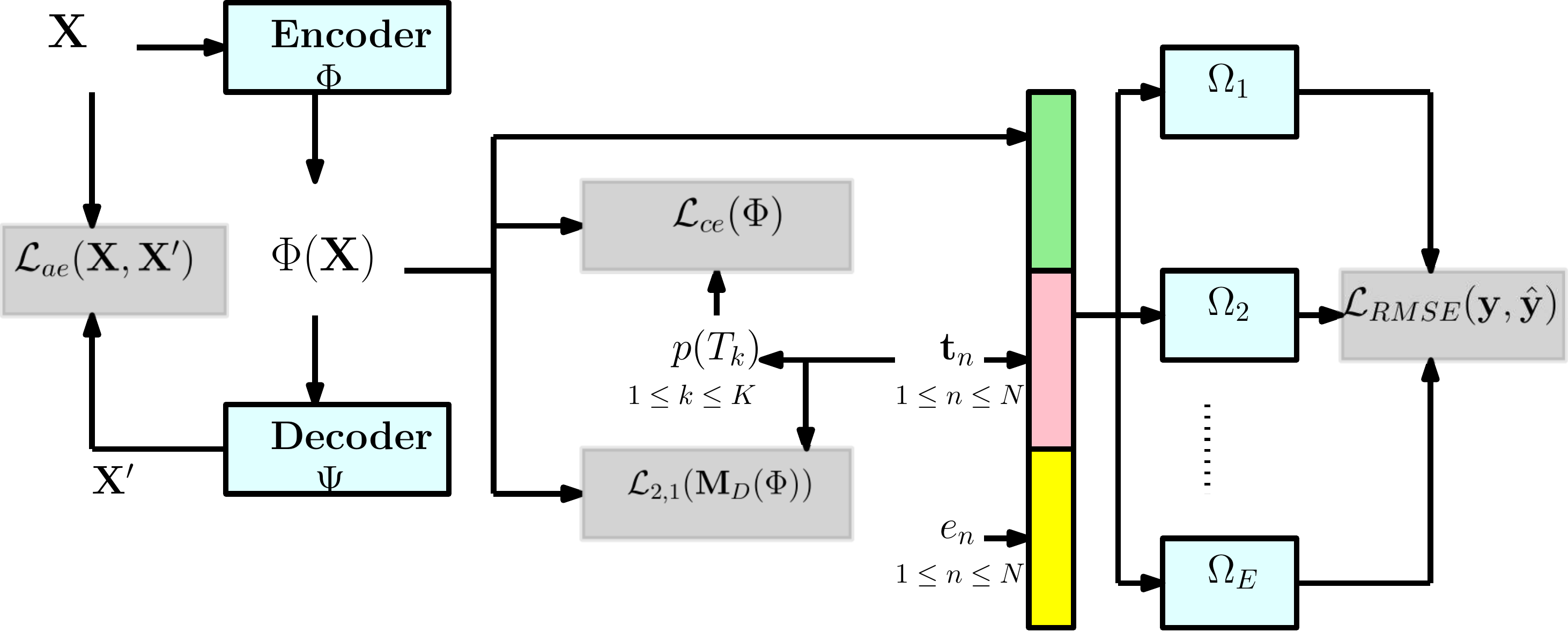}
\caption{\texttt{Hi-CI}: Proposed DNN architecture for inference in the presence of high-cardinality covariates and continuous treatments. The blocks in cyan represent the parameters to be tuned, and the blocks in grey denote the loss functions.}
\label{fig:HiCIDNN}
\end{figure*}

\section{Hi-CI: Proposed Deep Neural Network Architecture}
In this section, we describe the novel \texttt{Hi-CI} network architecture proposed for CI in high-dimensions. The network consists of two components, a decorrelation network and a regression (or outcome prediction) network. 
The decorrelation network consists of the autoencoder and the corresponding loss function is given in \eqref{eq:ellD}. The loss function pertaining to the regression network is given by \eqref{eq:lossRMSE}. The overall loss function corresponding to the decorrelation network and regression network is given by 
\begin{align}
    \mathcal{L}(\Phi,\Psi, \Omega, \beta , \gamma, \lambda) = \mathcal{L}_D(\Phi,\Psi,\beta,\gamma) 
     + \lambda \mathcal{L}_{RMSE}(\vecy,\hat{\vecy})
\label{eq:overallLoss}
\end{align}

As explained in the previous section, in the case of continuous treatments, we modify the structure of the regression network alone. In the regression network, we obtain the per-dosage level embedding, which we denote as $\Omega_e(\cdot)$, where $1 \leq e \leq E$. We use the concatenation of learned representation $\Phi(\vecx_n)$, treatment vector $\vect_n$ as an input to the embedding layer. The dosage information is used to obtain a subdivided network, i.e., in this work we split the network based on dosages and not treatments since $E \ll K$. The overall loss function for continuous treatments is given by 
\begin{align}
\mathcal{L}(\Phi, \Psi, \Omega_e, \beta,\gamma,\lambda) =  \mathcal{L}_D(\Phi,\Psi,\beta,\gamma) + \lambda\mathcal{L}_{RMSE}(\vecy,\hat{\vecy})
\label{eq:overallLoss_dose}
\end{align}
The generalized architecture of the \texttt{Hi-CI} network with continuous treatments is as depicted in Fig.~\ref{fig:HiCIDNN} (details in Supplementary Information). For discrete treatments, $E = 1$, and hence one embedding sub-network ($\Omega(\cdot)$) is used instead of multiple sub-networks ($\Omega_e(\cdot)$) for outcome prediction.

\section{Experimental Set-up}
In this section, we describe the experimental setup to demonstrate the efficacy in counterfactual regression of the \texttt{Hi-CI} algorithm and report the results obtained. We report the results on a synthetically generated dataset~\citep{sun2015causal}, and the semi-synthetic NEWS dataset~\citep{johansson2016learning} for evaluation.~\footnote{The simulated datasets will be available upon request from authors post publication of the paper.} Since a counterfactual outcome is not available, it becomes impossible to test CI algorithms in the context of counterfactual prediction. As a solution, data generating processes (DGP) are employed for demonstrating the results. In this section, we describe the datasets employed as well as the corresponding DGPs employed for each dataset. Furthermore, we describe the metrics we use for evaluating the \texttt{Hi-CI} framework where $E = 1$, namely precision in estimation of heterogeneous effect (PEHE)~\citep{pmlr-v70-shalit17a} and  Mean Absolute Percentage Error (MAPE) over Average Treatment Effect (ATE)~\citep{sharma2019metaci}. In the case of continuous treatments, i.e., for $E > 1$, we evaluate the \texttt{Hi-CI} framework using Mean Integrated Squared Error (MISE) and MAPE over ATE with dosage metric.

\subsection{Datasets}
In this subsection, we describe the datasets and the DGP employed for each dataset. 


\begin{itemize}
\item{Synthetic (\textsf{Syn})}: We use a synthetic process described in~\citep{sun2015causal} to generate data for both multiple treatment as well as continuous valued treatment scenario. The DGP gives us the flexibility to simulate the counterfactual responses along with the factual treatments and responses, thereby helping in better evaluation of the proposed model. The generation process in~\citep{sun2015causal} allows for $5$ confounding covariates while the remaining $P-5$ covariates are non-confounding. The number of covariates $P$, data size $N$ and cardinality of treatment set $K$ are fixed according to the requirement of experiment and is described in detail in section~\ref{sec:exptts}.

\item \textsf{NEWS}: We consider the publicly available bag-of-words context covariates for \textsf{NEWS} dataset \footnote{https://github.com/d909b/perfect\_match}. The DGP as given in~\citep{schwab2018perfect} is employed for synthesizing one of multiple treatments and corresponding response for each document (context) in \textsf{NEWS} dataset. This generation process is extended to treatments with dosage levels by~\citep{schwab2019learning} and is used for experimental evaluation of continuous valued treatments. The number of covariates $P$ is fixed to $2870$ and value for $N,K$ is given in sec.~\ref{sec:exptts} based on experimental requirements.
\end{itemize}
We use the convention of naming each newly synthesized dataset as a conjunction of the original dataset name and the treatment set cardinality ($K$) for all experiments in section~\ref{sec:exptts}.  For example, '\textsf{NEWS4}' denotes \textsf{NEWS} dataset for $K=4$ treatment case.

\subsection{Metrics}
In this subsection, we describe the performance metrics used for evaluating the proposed approach.
\subsubsection{Precision in Estimation of Heterogeneous Effect (PEHE)} 
We use the definition of PEHE as specified in \citep{schwab2018perfect} for multiple treatments as:
\begin{equation}
    {\hat\epsilon_{P}} = \frac{1}{\binom{K}{2}}\sum_{m=1}^{K}\sum_{r=1}^{m-1}{\hat\epsilon_{P_{m,r}}}
\end{equation}
\begin{equation}
    \hat\epsilon_{P_{m,r}} = \frac{1}{N}\sum_{n=1}^{N}([y_{n}(m) - y_{n}(r)] - [\hat y_{n}(m) - \hat y_{n}(r)])^2
\end{equation}
where $y_{n}(m)$,$y_{n}(r)$ are the response of the $n$-{th} individual to treatments $T_m$ and $T_r$ respectively.
\subsubsection{Mean Absolute Percentage Error (MAPE) over Average Treatment Effect (ATE)}
We use $MAPE_{ATE}$ as a metric to estimate error in predicting average treatment effect for high-cardinality treatments, and is given by:
\begin{equation}
    MAPE_{ATE} = \left|\frac{ATE_{actual} - ATE_{pred}}{ATE_{actual}}\right|,
\end{equation}
where,
\begin{equation}
    ATE_{actual} = \frac{1}{N}\sum_{n=1}^{N} \left(y_{n}(k) - \frac{1}{K-1}\sum_{l=1, l \neq k}^{K}y_{n}(l)\right),
\end{equation}
and $ATE_{pred}$ is obtained by replacing $y_{n}(k)$ in the above equation by its predicted value $\hat y_{n}(k)$ for all $k$.

\subsubsection{Mean Integrated Squared Error (MISE)} For high cardinality treatments with dosages, we use $MISE$ as a metric (as in~\citep{schwab2019learning}). This is the squared error of dosage-response computed across the dosage levels and averaged over all treatments and entire population.

\subsubsection{MAPE over ATE with dosage} We introduce a novel metric $MAPE_{ATE}^{Dos}$ for high cardinality treatments with dosages. This metric is useful for evaluating effect of a dosage level for factual treatment as opposed to counterfactual treatments.
It is given by:
\begin{equation}
    MAPE_{ATE}^{Dos} = \left|\frac{ATE_{actual}^{Dos} - ATE_{pred}^{Dos}}{ATE_{actual}^{Dos}}\right|,
\end{equation}
where,
\begin{equation}
    ATE_{actual}^{Dos} = \frac{1}{E}\sum_{e=1}^{E}\left(\frac{1}{N_{E}}\sum_{n=1}^{N_{E}}\left(y_{n}(k_e) - \frac{1}{K-1}\sum_{l=1, l \neq k}^{K}y_{n}(l_e)\right)\right),
\end{equation}
and and $ATE_{pred}^{Dos}$ is obtained by replacing $y_{n}(k_e)$ in the above equation by its predicted value $\hat{y}_{n}(k_e)$ for all $k$.



\subsection{Baselines}
\label{sec:baselines}
 We use the following DNN based approaches to baseline Hi-CI for high cardinality treatments: 

\begin{itemize}
    \item \texttt{O-NN}: O-NN does not account for confounding bias, we thereby bypass the decorrelation network of \texttt{Hi-CI} and $\matX$ is directly passed to the outcome network.
    \item \texttt{MultiMBNN}: Matching and balancing based architecture proposed in \citep{esann-multiMBNN}
    \item \texttt{PM} : Propensity based matching \citep{schwab2018perfect} employed for counterfactual regression.
    \item \texttt{Deep-Treat+}: Deep-Treat \citep{atan2018deep} learns bias removing network and policy optimization network independently to learn optimal, personalized treatments from observation data. In order to use Deep-treat as a baseline, we modify Deep-Treat to \texttt{Deep-Treat+} and jointly train decorrelation network obtained from Deep-Treat, and outcome network of \texttt{Hi-CI} to baseline our approach.
    \item \texttt{DRNet}: is a DNN based technique \citep{schwab2019learning} to infer counterfactual responses when treatments have dosage values. We use this to baseline \texttt{Hi-CI} continuous valued treatment case.
\end{itemize}




\section{Experimental Results}
\label{sec:exptts}
We perform extensive experimentation using the proposed \texttt{Hi-CI} framework on \textsf{Syn} and \textsf{NEWS} datasets. Our experimental evaluation is primarily aimed at evaluating the performance of \texttt{Hi-CI} under three broad settings: high-cardinality treatments; continuous valued treatments and high number of covariates.

\subsection{High-cardinality treatments ($E=1$)}

\subsubsection{Effect of increasing the cardinality of treatment set}
Here, we evaluate \texttt{Hi-CI} in scenarios where the cardinality of treatments increases, while $E = 1$ . With increase in $K$, sample size $N$ is also proportionally increased to keep the average number of samples per treatment (given by $\frac{N}{K}$) constant. In Table~\ref{tab:exp2_fixed_n_k_ratio}, we report the mean and standard deviation of the performance metrics $PEHE, MAPE_{ATE}$ for \textsf{Syn} and \textsf{NEWS} datasets. For both the datasets, performance errors increase with increase in $K$. In the case of \textsf{Syn} dataset, error in estimating ATE is much lower than \textsf{NEWS} dataset for very large number of treatments. We can explain this by noting that the number of covariates (perhaps confounding too) in  \textsf{NEWS} dataset are of the order of $2000$ whereas in \textsf{Syn}, the number of covariates are fixed to $10$ with $5$ confounding variables.
\begin{table}
\centering
\caption{Effect of increasing $K$ with fixed $\tfrac{N}{K}$ ratio ($\tfrac{N}{K}$ = 285.7) on high cardinality treatments with $E=1$. Mean and standard deviation of the metrics for multiple runs with varying seed points is reported.}
\label{tab:exp2_fixed_n_k_ratio}  
  \begin{tabular}{l|l|l}
    \hline
    Dataset & $\sqrt{\hat\epsilon_{P}}$ & $MAPE_{ATE}$\\
    \hline
    \textsf{Syn35} & 3.6764, 0.4037 & 0.074, 0.0188\\
    \textsf{Syn48} & 7.4350, 0.1705 & 0.1494, 0.0048\\
    \textsf{Syn103} & 7.0612, 0.5124 & 0.1681, 0.0054\\
    \textsf{Syn216} & 7.7069, 0.1531 & 0.1943, 0.0101\\
    \textsf{NEWS35} & 7.6256, 0.0243 & 0.393, 0.0095\\
    \textsf{NEWS48} & 8.2675, 0.0522 & 0.4821, 0.0105\\
    \textsf{NEWS100} & 8.9334, 0.6425 & 0.566, 0.0245\\
    \textsf{NEWS200} & 9.4679, 0.8524 & 0.8924, 0.0859\\
    \hline
\end{tabular}
\end{table}

\subsubsection{Varying number of treatments $K$ for fixed $N$}
\begin{figure}
\centering
\includegraphics[width=0.55\linewidth]{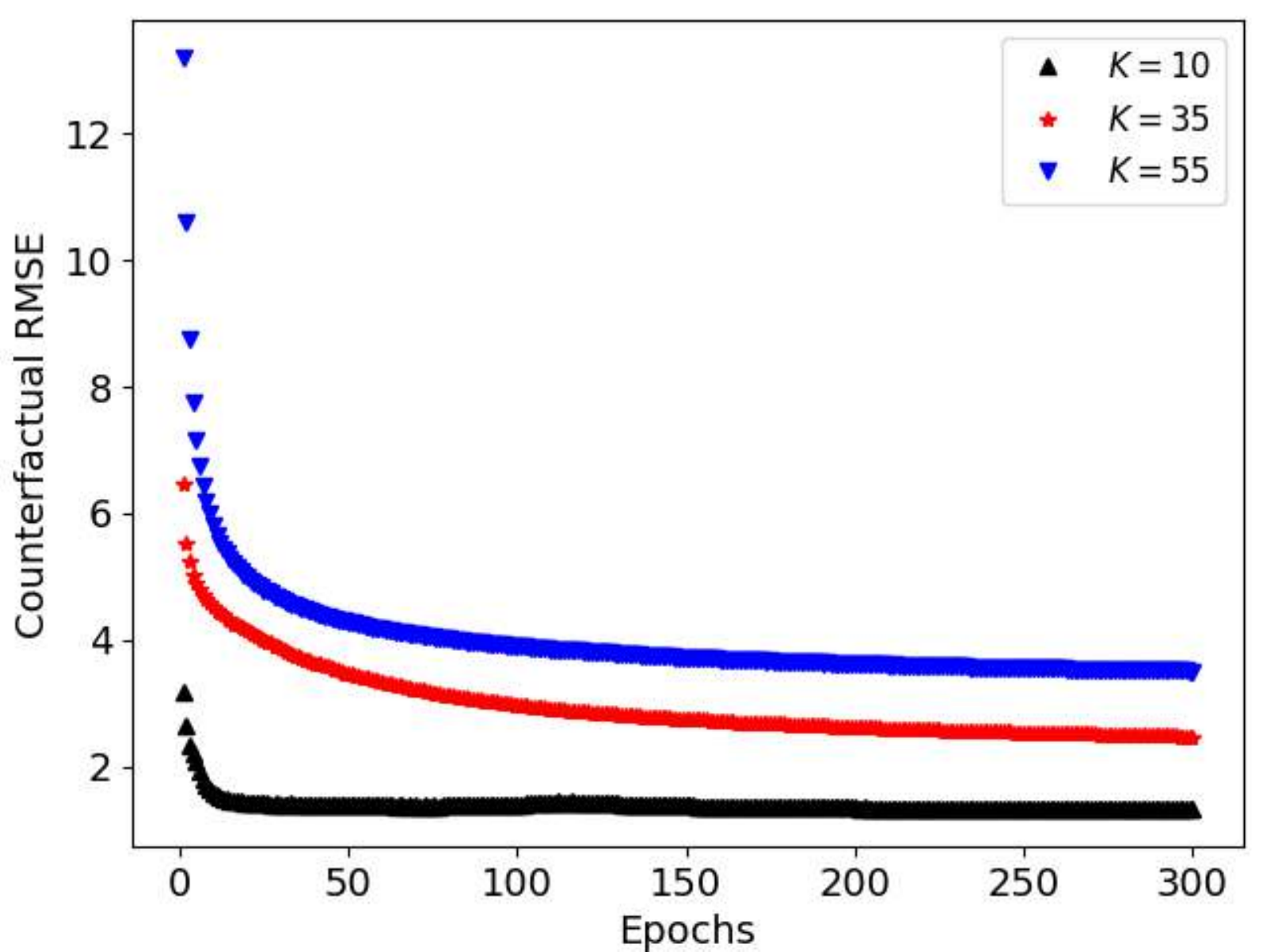}
\caption{Counterfactual RMSE for increasing $K$, constant $N=10000$ on \textsf{Syn$K$} datasets.} 
\label{fig:increaseK}
\end{figure}

We illustrate the performance of the \texttt{Hi-CI} framework keeping a sample size of $N=10000$ while we vary the cardinality of treatment set from $K=10$ to $100$, which implies that we see a decrease in the ratio  $\frac{N}{K}$. From Table~\ref{tab:varying NbyK}, we observe that for $Syn$ dataset, as the average number of samples per treatment decreases, $PEHE$ and $MAPE_{ATE}$ increase. However, for the \textsf{NEWS} dataset, no such trend is observed due to a large number but sparse covariates. Furthermore, in Fig.~\ref{fig:increaseK} we depict the counterfactual RMSE for \textsf{Syn} datasets under this experimental setting. We observe a slight increase in the  counterfactual error as $K$ increases,  demonstrating that although the problem is harder, \texttt{Hi-CI} network prediction performs reasonably well.

\begin{table}[]
\centering
\caption{Mean and standard deviation of $\sqrt{\hat\epsilon_{P}}$ , $MAPE_{ATE}$ for \textsf{NEWS , Syn} datasets with varying $\frac{N}{K}$ ratios. Results are computed for multiple runs of each dataset simulated for varying seed points.}
    \label{tab:varying NbyK}
    \begin{tabular}{l|l|l|l}
        \hline
        Dataset & $\frac{N}{K}$ & $\sqrt{\hat\epsilon_{P}}$ & $MAPE_{ATE}$\\
        \hline
        \textsf{Syn10} & 1000 & 1.6188,0.0262 & 0.046,0.008\\
        \textsf{Syn35} & 285.7 & 3.6764,0.4037 & 0.074,0.0188\\
        \textsf{Syn55} & 181.8 & 7.1836,0.8065 & 0.1378,0.0148\\
        \textsf{Syn100} & 100 & 9.1706,0.8755 & 0.1812,0.0129\\
        \textsf{NEWS10} & 1000 & 7.8563,0.0214 & 0.6223,0.115\\
        \textsf{NEWS35} & 285.7 & 7.6256,0.0243 & 0.393,0.0095\\
        \textsf{NEWS55} & 181.8 & 7.7383,0.0273 & 0.4515,0.0360\\
        \textsf{NEWS100} & 100 & 8.1432,0.0476 & 0.507,0.0171\\
        \hline
    \end{tabular}
\end{table}

\subsubsection{Loss Functions Analysis}
\begin{table}
\centering
\caption{Effect of decorrelation loss function $\mathcal{L}_D(\cdot)$ as compared to $\mathcal{L}_{ce}(\cdot) + \mathcal{L}_{ae}(\cdot)$ and $\mathcal{L}_{2,1}(\cdot) + \mathcal{L}_{ae}(\cdot)$ for $E=1$ case.} 

\label{tab:exp1_high_card_treat}
\resizebox{\textwidth}{!}{
\begin{tabular}{lll|lll|lll}
\hline
& & &  &$\sqrt{\hat\epsilon_{P}}$ & &  & $MAPE_{ATE}$& \\
\hline
Dataset & 
$\tfrac{N}{K}$ & $P$ & $\mathcal{L}_{1}(\cdot)$ & $\mathcal{L}_{2,1}(\cdot) + \mathcal{L}_{ae}(\cdot)$ & $\mathcal{L}_D(\cdot)$ & $\mathcal{L}_{1}(\cdot)$ & $\mathcal{L}_{2,1}(\cdot) + \mathcal{L}_{ae}(\cdot)$ & $\mathcal{L}_D(\cdot)$\\
\hline
\textsf{Syn10} & 1000 & 10 & 1.6390, 0.1125 & \textbf{1.6161, 0.0506} & 1.6188, 0.0262 & 0.0645, 0.0243 & 0.0573, 0.0111 & \textbf{0.046, 0.008}\\
\textsf{Syn35} & 285.7 & 10  & 5.3784, 0.4538 & 4.2283, 0.8902 & \textbf{3.6764, 0.4037} & 0.1686, 0.0181 & 0.0990, 0.0407 & \textbf{0.074, 0.0188} \\
\textsf{Syn55} & 181.8 & 10 & 7.5039, 0.4699 & 7.5173, 0.5540 & \textbf{7.1836, 0.8065} & 0.1443, 0.0115 & 0.1472, 0.0103 & \textbf{0.1378, 0.0148}\\
\textsf{Syn100} & 100 & 10 & 9.7575, 0.7000 & 11.3353, 0.7624 & \textbf{9.1706, 0.8755} & 0.2214, 0.0394 & 0.2138, 0.0067 & \textbf{0.1812, 0.0129}\\
\textsf{NEWS10} & 1000 & 2870 & 7.8601, 0.0487 & \textbf{7.8541, 0.0285} & 7.8563, 0.0214 & 0.6288, 0.0146 & 0.6325, 0.0027 & \textbf{0.6223, 0.115}\\
\textsf{NEWS35} & 285.7 & 2870 & 8.3121, 0.0442 & 8.3425, 0.0600 & \textbf{7.6256, 0.0243} & 0.4875, 0.0141 & 0.4874, 0.0081 & \textbf{0.393, 0.0095}\\
\textsf{NEWS55} & 181.8 & 2870 & 7.8019, 0.0297 & 7.8212, 0.1648 & \textbf{7.7383, 0.0273} & 0.4792, 0.0173 & 0.6454, 0.0945 & \textbf{0.4515, 0.0360}\\
\textsf{NEWS100} & 100 & 2870 & 8.3275, 0.0792 & 8.2897, 0.0284 & \textbf{8.1432, 0.0476} & 0.5028, 0.0169 & \textbf{0.4844, 0.0050} & 0.507, 0.0171\\
\hline
\end{tabular}}
\end{table}
We conducted extensive experimentation to validate the impact of the proposed decorrelation loss function $\mathcal{L}_{D}(\cdot)$ given in \eqref{eq:ellD}, in learning the low-dimensional representation of data as the cardinality of treatments increases. We set the sample size to be  constant while $K$ increases, and consequently the ratio $\frac{N}{K}$ decreases. From Table~\ref{tab:exp1_high_card_treat}, we see that $PEHE$  and $MAPE_{ATE}$) decrease significantly when the lower-dimensional representation is learned using $\mathcal{L}_D(\cdot)$ loss function \eqref{eq:ellD}, a combination of losses that caters to reduction in bias via $\mathcal{L}_{ce}(\cdot)$, reduction in information loss via $\mathcal{L}_{ae}(\cdot)$, and similarity-exploiting via $\mathcal{L}_{2,1}(\cdot)$ as compared to case where only $\mathcal{L}_{1}(\cdot)$ or $\mathcal{L}_{ae}(\cdot) + \mathcal{L}_{2,1}(\cdot)$ is used. Note that $\mathcal{L}_{1}(\cdot)$ is considered as decorrelation loss in Deep-Treat+.

\subsection{Varying number of covariates $P$}
We illustrate the performance of the \texttt{Hi-CI} framework by increasing the number of covariates, retaining the sample size fixed at $N=10000$, i.e., $\frac{P}{N}$ varies from $0.001$ to $0.1$. In the context of \textsf{Syn35} dataset, we see from Table~ \ref{tab:exp3_noisy_covariates} that as the number of covariates increase, $\sqrt{\hat\epsilon_{P}}$ is as low as $3.67$ and $MAPE_{ATE}$ is lower than $0.17$, thereby showing the strength of the proposed network in handling high dimensional covariates.
\begin{table}[h]
\centering
\caption{Effect of increasing $\frac{P}{N}$ for \textsf{Syn35} dataset on $PEHE$ ($\sqrt{\hat\epsilon_{P}}$) and error in ATE measured using $MAPE_{ATE}$.}
 \label{tab:exp3_noisy_covariates}
  \begin{tabular}{l|l|l}
    \hline
    $\frac{P}{N}$ & $\sqrt{\hat\epsilon_{P}}$ & $MAPE_{ATE}$\\
    \hline
    0.001 & 3.6764, 0.4037 & 0.074, 0.0188\\ 
    0.005 & 5.1845, 0.7025 & 0.1388, 0.0192\\ 
    0.01 & 6.2392, 0.3310 & 0.1557, 0.0132\\ 
    0.05 & 6.0466, 0.4325 & 0.1720, 0.0104\\ 
    0.1 & 6.2516, 0.6775 & 0.1757, 0.0260\\ 
    \hline
\end{tabular}
\end{table}

\subsection{High-cardinality treatments with continuous dosages ($E>1$)}
In Table~\ref{tab:exp3_varying_e}, we illustrate the effect of varying number of dosage levels on the performance metrics for treatments with dosage. Note that the error decreases as the number of dosage levels $E$ increase. We measure the dose-reponse error using $MISE$ and average dosage effect given by $MAPE_{ATE}^{Dos}$ in Table~\ref{tab:exp3_varying_e} show that varying dosage levels does not impact the performance much. Note that this is partly since context covariates are confounders for treatments, but not for dosage levels in the $NEWS$ dataset. Furthermore, In case of synthetic dataset, although covariates are confounders for both treatments and dosages, we see that low-complexity networks are sufficient to capture the dosage-response. As pointed in Sec.~\ref{sec:prelim} that our network is designed under the assumption treatment is confounded but not dosage values. However the results for \textsf{Syn} dataset, as seen in Table~\ref{tab:exp3_varying_e}, show that \texttt{Hi-CI} can handle covariates confounding dosages as well.
\begin{table}[h]
\centering
\caption{Varying dosage levels $E$ for $N = 10,000$ on treatments with dosage. Mean and standard deviation of $\sqrt{MISE}$, $MAPE_{ATE}^{Dos}$ computed on test set of datasets generated by varying seeds.}
  \label{tab:exp3_varying_e}  
  \begin{tabular}{l|l|l|l}
    \hline
    Dataset & $E$ & $\sqrt{MISE}$ & $MAPE_{ATE}^{Dos}$\\
    \hline
    \textsf{Syn25} & 3 & 2.126, 0.0146 & 0.1193, 0.0024\\
    \textsf{Syn25} & 6 & \textbf{1.980, 0.0157} & \textbf{0.1066, 0.0038}\\
    \textsf{Syn25} & 8 & 2.146, 0.014 & 0.124, 0.0021\\
    \textsf{Syn25} & 10 & 3.148, 0.052 & 0.162, 0.0046\\
    \textsf{NEWS25} & 3 & 11.2346, 0.1221 & 0.2462, 0.0584\\
    \textsf{NEWS25} & 6 & 11.4860, 0.1568 & 0.3254, 0.1221\\
    \textsf{NEWS25} & 8 & \textbf{11.0114, 0.0856} & \textbf{0.1457, 0.0462}\\
    \textsf{NEWS25} & 10 & 11.9086, 0.2795 & 0.6890, 0.1258\\
    \hline
\end{tabular}
\end{table}

\subsection{Comparative analysis with baselines}
In this section, we illustrate the performance of the \texttt{Hi-CI} network as compared to the popular baselines in literature. 

\subsubsection{High dimension treatments and covariates for $E=1$}
In Table~\ref{tab:baseline_E=1}, we depict the performance of \texttt{Hi-CI} framework as compared to baselines (refer Section \ref{sec:baselines}) with varying number of treatments for low and high dimensional covariates. In order to evaluate the performance in high dimensions, \textsf{NEWS100} with $\frac{P}{N} = 0.287$ is shown to do exceedingly well in terms of both $\sqrt{\hat\epsilon_{P}}$ and $MAPE_{ATE}$, as compared to previous works. It is seen that for lower-cardinality treatment set (\textsf{Syn4, NEWS4}) our approach beats state of art marginally. This is expected behavior since baselines such as \citep{esann-multiMBNN} and \citep{schwab2018perfect} are optimized for such scenarios. However as the number of treatments increase, our approach outperforms baselines by huge margins. This behaviour is observed for both high and low number of covariates.

\begin{figure}[h!]
\centering
\includegraphics[scale=0.70]{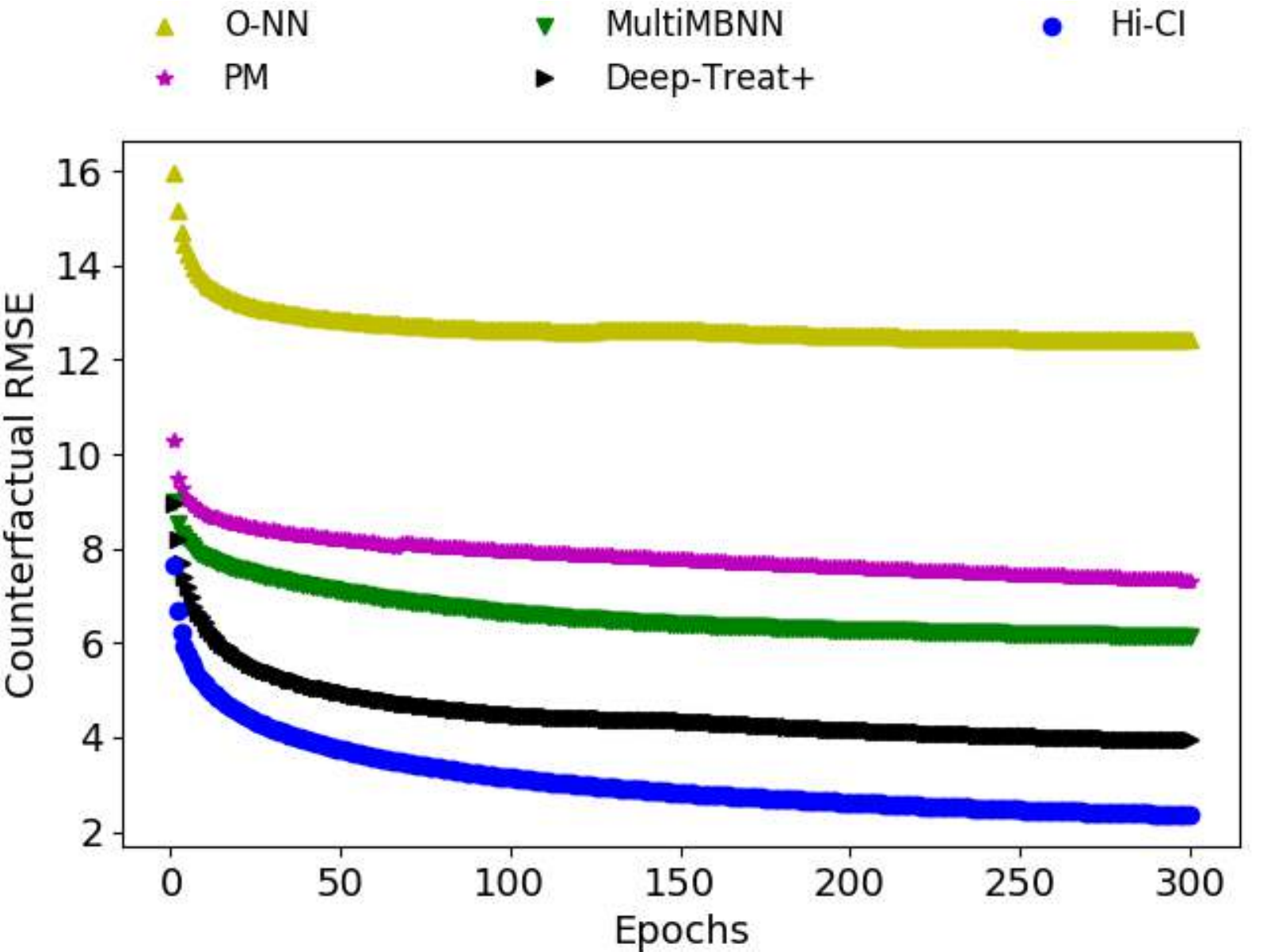}
\caption{Comparison against baselines using counterfactual RMSE metric on \textsf{Syn10} dataset for $N=10000$.}
\label{fig:Baselines}
\end{figure}

In Fig.~\ref{fig:Baselines}, we depict the counterfactual RMSE obtained using \texttt{Hi-CI} as compared to \texttt{O-NN}, \texttt{PM}, \texttt{Deep-Treat+}, validating our claim of outperforming the state of art approaches for CI.

\begin{table}[]
\centering
\caption{Performance of \texttt{Hi-CI} as compared to state-of-the-art approaches: \texttt{PM}, \texttt{MultiMBNN} for high cardinality treatments.}
    \label{tab:baseline_E=1}
    \resizebox{\textwidth}{!}{
    \begin{tabular}{lll|lll|lll}
    \hline
    & & & &$\sqrt{\hat\epsilon_{P}}$ & & &$MAPE_{ATE}$&\\
    \hline
    Dataset & $\frac{P}{N}$ & $\frac{N}{K}$ & \texttt{PM} & \texttt{MultiMBNN} & \texttt{Hi-CI} & \texttt{PM} & \texttt{MultiMBNN} & \texttt{Hi-CI}\\
         \hline
         \textsf{Syn4} & 0.001 & 2500 &  1.9004, 0.1124 & 1.8272, 0.0928 & \textbf{1.3520, 0.0542} & 0.4249, 0.1142 & 0.3917, 0.1075 & \textbf{0.0150, 0.0022}\\
         \textsf{Syn10} & 0.001 & 1000 & 0.4249, 0.1142 & 0.3917, 0.1075 & \textbf{0.0150, 0.0022} & 5.8976, 0.1175 & 5.7752, 0.1100 & \textbf{1.6188, 0.0262}\\
         \textsf{Syn35} & 0.001 & 285.7 & 18.5894, 0.2329 & 17.6520, 0.2032 & \textbf{3.6764, 0.4037} & 0.4726, 0.0562 & 0.4528, 0.0864 & \textbf{0.074, 0.0188}\\
         \textsf{Syn100} & 0.001 & 100 & 32.0424, 0.9862 & 32.304, 0.9652 & \textbf{9.1706, 0.8755} & 1.1225, 0.1585 & 1.2854, 0.2012 & \textbf{0.1812, 0.0129}\\
         \textsf{NEWS4} & 0.287 &2500 &8.1842, 0.4202 & 7.6606, 0.4077 & \textbf{6.4120, 0.3016} & 0.3232, 0.0574 & 0.1622, 0.0381 & \textbf{0.0984, 0.0245}\\
         \textsf{NEWS10} & 0.287 & 1000 & 9.1540, 0.0245 & 9.002, 0.0185 & \textbf{7.8563, 0.0214}& 0.8641, 0.0962 & 0.7452, 0.105 & \textbf{0.6223, 0.115}\\
         \textsf{NEWS35} & 0.287 & 285.7 & 18.5894, 0.2329 & 17.6520, 0.2032 & \textbf{3.6764, 0.4037} & 0.4726, 0.0562 & 0.4528, 0.0864 & \textbf{0.074, 0.0188}\\
         \textsf{NEWS100} & 0.287 & 100 & 48.3878, 0.5620 & 49.6386, 0.8520 & \textbf{8.1432, 0.0476} & 1.9850, 0.1824 & 2.2014, 0.2350 & \textbf{0.507, 0.0171}\\
         \hline
    \end{tabular}}
\end{table}

\subsubsection{High cardinality treatments with continuous dosages}
In Table~\ref{tab:baselines_high_card_treat_dosages} we depict the comparative dosage-response values for different datasets averaged over all treatments and individuals, in terms of $\sqrt{MISE}$. We observe that the \texttt{Hi-CI} framework outperforms the state of the art DNN-based approach, \texttt{DRNet} by a considerable margin for several treatment counts.
\begin{table}
\centering
\caption{Comparing with baselines: \texttt{Hi-CI} for continuous treatments, $E>1$.}
  \label{tab:baselines_high_card_treat_dosages}
  \begin{tabular}{l|l|l}
    \hline
    Dataset & \texttt{DRNet} & \texttt{Hi-CI}\\
    \hline
    \textsf{NEWS2} & 7.7, 0.2 & \textbf{6.2450, 0.1254}\\
    \textsf{NEWS4} & 11.5, 0.0 & \textbf{11.0842, 0.1358}\\
    \textsf{NEWS8} & 10.0, 0.0 & \textbf{8.7540, 0.1032}\\
    \textsf{NEWS16} & 10.2, 0.0 & \textbf{8.6560, 0.0452}\\
  \hline
\end{tabular}
\end{table}

\section{Discussions and Conclusions}

In CI applications, one commonly encounters situations where there are large number of covariates and large number of treatments in real-world observational studies. The biggest hindrance in such a scenario is in inferring which of the covariates is the actual confounder among the large number of covariates. Furthermore, the complexity of the situation is enhanced since one needs to determine such confounding effects per treatment, for a large number of treatments. 

We tackle these seemingly hard scenarios using a generalized \texttt{Hi-CI} framework. Our approach is based on a fundamental assumption that the high-dimensional covariates are often sparse, and can be represented in a low-dimensional space. Naturally, we employed an auto-encoder to represent covariates in a low-dimensional space, without losing much information in the original covariates. Alongside, we also incorporate a decorrelating loss function which ensures that we obtain an equivalent representation of the covariate space with a reduced confounding bias. Furthermore, using the fact that often several treatments/interventions are perhaps similar, we used an embedding in order to obtain a low-dimensional representation of the treatment. In literature, continuous treatments are used, which we addressed by using per-dosage level embedding. 

We performed extensive experimental analysis of \texttt{Hi-CI} for synthetic \textsf{Syn} dataset and semi-synthetic \textsf{NEWS} dataset. For high cardinality discrete treatments, we observed that when we increase the  number of treatments, not only does \texttt{Hi-CI} perform considerably well in spite of CI being an ill-posed problem, but also is able to outperform the state of art DNN approaches. We further show that the proposed framework handles continuous valued dosages which may or not may be confounded by covariates. We also depicted the decreasing trend in RMSE across epochs, which leads us to understand that the model has indeed trained well. Overall, the theory and results demonstrate that \texttt{Hi-CI} can be regarded as a generalized causal inference framework for tackling high dimension covariates and large number of treatments which may be continuous or discrete valued in observational studies. As a future work, we intend to analyse scenarios where $P \ll N$, which occur in biological applications such as gene intervention studies.


\newpage

\appendix
\section{}
In this appendix, we provide details of the experiments performed to evaluate \texttt{Hi-CI} framework. Specifically, we describe the dataset preparation, training mechanism and hyperparameter tuning.

\subsection{Implementation}

\subsubsection{Algorithm}
We provide the implementation details of \texttt{Hi-CI} framework in Algorithm \ref{alg:HiCI} and \ref{alg:trainer}. Specifically, in algorithm \ref{alg:HiCI} we explain the methodology used for splitting dataset $\matD$ into train ($\matD_{CI}$), validation ($\matD_{val}$), test ($\matD_{tst}$) sets. We also explain the mechanism for hyperparameter selection. On the otherhand, algorithm \ref{alg:trainer} outlines the procedure for training \texttt{Hi-CI} for the given set of hyperparameters. We initialise the parameters $W$ of \texttt{Hi-CI} using random normal distribution. We use Adam optimizer with inverse time decay learning rate for gradient descent. In algorithm \ref{alg:HiCI},  $hparam\_values$ specifies the range of hyperparameters for grid-search as in Table~\ref{tab:param_value_hici}, $num\_unique\_treat(\cdot)$ returns the number of unique treatments in the dataset passed as argument, $get\_gs\_hparams(\cdot)$ returns set containing exhaustive combination of hyperparameters,  $get\_best\_params(\cdot)$ returns \texttt{Hi-CI} parameters corresponding to best validation loss and $get\_metric(\cdot)$  returns performance metrics of trained \texttt{Hi-CI} on dataset passed as argument. Similarly in algorithm \ref{alg:trainer}, $initialize(\cdot)$ initializes parameters of \texttt{Hi-CI} using random normal distribution,  $get\_random\_batches(\cdot)$ creates random batches of the dataset with batch size as specified in the argument, $train(\cdot)$ trains \texttt{Hi-CI}, $check\_convergence(\cdot)$ checks for convergence on $\matD_{val}$, 
$get\_final\_params(\cdot)$ returns learned parameters $W_f$ of \texttt{Hi-CI} and $get\_val\_loss(\cdot)$ returns loss $\mathcal{L}$ on $\matD_{val}$ corresponding to $W_f$.


\begin{algorithm}
 \caption{\texttt{Hi-CI}} 
\label{alg:HiCI}
\begin{algorithmic}[1]
\Procedure{\texttt{Hi-CI}}{$\matD$, $K$, $hparam\_values$, $E=1$}
\State Split $\matD$ into $\matD_{CI}$,$\matD_{val}$,$\matD_{tst}$ 
\While {$num\_unique\_treat(\matD_{CI}) < K$}
    \State Split $\matD$ into $\matD_{CI}$,$\matD_{val}$,$\matD_{tst}$
\EndWhile
\State $\mathcal{L}_{val} = \emptyset, W = \emptyset$
\State $gs\_hparams = get\_gs\_hparams(hparam\_values)$\
\For{$gs\_hparam$ in $gs\_hparams$}
    \State $\mathcal{L}_{val}, W \gets trainer(gs\_hparam, \matD_{CI}, \matD_{val})$
\EndFor
\State $W' = get\_best\_params(\mathcal{L}_{val}, W)$
\State $PEHE, MAPE_{ATE} = get\_metric(\matD_{tst}, W')$ \par
\Return $PEHE, MAPE_{ATE}, W'$
\EndProcedure
\end{algorithmic}
\end{algorithm}

\begin{algorithm}
\caption{\texttt{Train$_{Hi-CI}$}}
\label{alg:trainer}
\begin{algorithmic}[1]
\Procedure{$trainer$}{$gs\_hparam, \matD_{CI}, \matD_{val}$}
\State $W = initialize()$
\State $total\_epochs = gs\_hparam.total\_epochs$
\State $batch\_size = gs\_pharam.batch\_size$
\While{$epoch <= total\_epochs$}
    \State $\matD_{batches} = get\_random\_batches(\matD_{CI}, batch\_size)$
    \For{$\matD_{batch}$ in $\matD_{batches}$}
        \State $W = train(W, \matD_{batch}, gs\_hparam)$
    \EndFor
    \If{$check\_convergence(W,\matD_{val})$}
    \State $W_f = get\_final\_params(W)$ 
    \State $break$
    \EndIf
    \State $epoch = epoch +1$
\EndWhile
\State $\mathcal{L} = get\_val\_loss(W_f, \matD_{val})$\par
\Return $\mathcal{L},W_f$
\EndProcedure
\end{algorithmic}
\end{algorithm}

\subsubsection{Parameter Tuning and Model Selection}
We select the optimal parameters $W'$ for \texttt{Hi-CI} by performing an exhaustive grid-search on the hyperparameters values mentioned in Table ~\ref{tab:param_value_hici}.

\begin{table}
  \centering
  \caption{Hyperparameters for grid-search ($hparam\_values$)}
  \label{tab:param_value_hici}
    \begin{tabular}{l|l}
    \hline
    \textbf{Parameter} & \textbf{Values}\\ \hline
    Batch size & 64, 128, 256, 512\\
    Total epochs & 1000\\
    Learning rate & 0.06, 0.08, 0.1, 0.12, 0.14, 0.16\\
    Learning rate decay & 0.6, 0.65, 0.7, 0.75\\
    No. of iterations per decay & 1, 2\\
    Train set split ratio & 0.6\\
    Validation set split ratio & 0.2\\
    Test set split ratio & 0.2\\
    No. of encoder layers  & 1, 2, 3, 5, 7\\
    No. of decoder layers  & 3, 4, 5, 6, 7, 8 \\
    No. of outcome layers  & 3, 4, 5, 6, 7, 8\\
    No. of hidden nodes in encoder layers & 100, 150, 200, 250\\
    No. of hidden nodes in decoder layers & 100, 175, 250, 325, 400\\
    No. of hidden nodes in outcome network & 100, 200, 250, 300, 400, 500\\
    L-2 regularization co-efficient for $\Phi,\Psi,\Omega_e$ & 0.01, 0.001, 0.0001\\ \hline
 \end{tabular}
\end{table}

\subsubsection{Learning $\theta_{T_{k}}$:}
We use the multi-class logistic regression library of scikit-learn for learning $\theta_{T_{k}}$ in \eqref{eq:logisticloss}. The range of hyperparameters for grid-search in logistic regression is given in Table~\ref{tab:param_value_lr}.

\begin{table}
  \centering
  \caption{Logistic regression hyperparameters for grid-search}
  \label{tab:param_value_lr}
  \begin{tabular}{l|l}
    \hline
    \textbf{Parameter} & \textbf{Values}\\ \hline
    Inverse of regularization strength & 0.001, 0.01, 0.1, 1, 10 \\
    Solver & newton-cg, sag, saga, lbfgs\\ 
    Tolerance for stopping criteria & 1e-4, 1e-2\\ \hline
\end{tabular}
\end{table}

\subsection{Baselines}
In order to make fair comparison of \texttt{Hi-CI} with baselines, we used consistent train, validation and test sets. Additionally, same set of hyperparameters were used for best model selection across all approaches.

\subsection{Compute Infrastructure}
All experiments in this paper were run on Nvidia Tesla V100 GPU with 32 GB memory and Intel Xeon CPUs with 64 GB RAM.

\begin{figure*}[h]
\includegraphics[width=\textwidth]{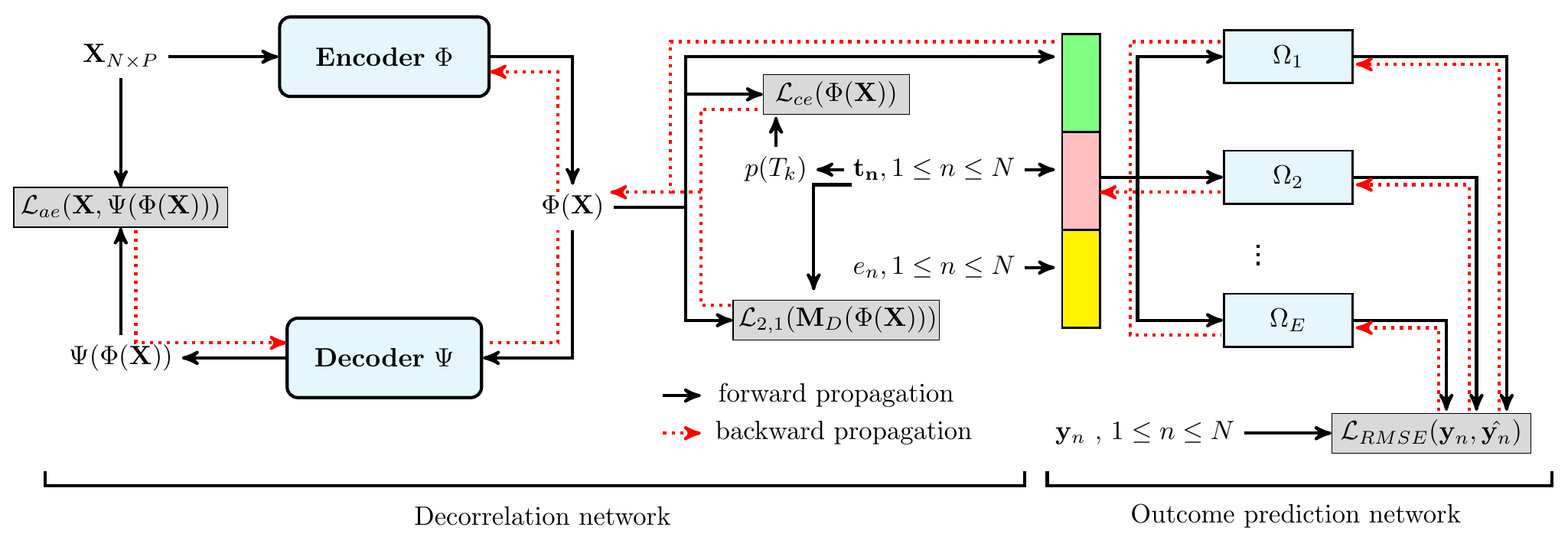}
\caption{Detailed architecture (with notations) of \texttt{Hi-CI} is depicted. Dotted red arrows highlight joint learning of decorrelating and outcome prediction network. $\mathbf{X} \in \mathbb{R}^{N \times P}$, $\Psi(\Phi(\mathbf{X})) \in \mathbb{R}^{N \times P}$, $\Phi(\mathbf(X)) \in \mathbb{R}^{N \times L}$, $\mathbf{t}_n \in \mathbb{R}^{K \times 1}$, $1 \leq k \leq K$. }
\label{fig:AppendixArch}
\end{figure*}
\vskip 0.2in
\bibliography{acmart}

\begin{thebibliography}{27}
\providecommand{\natexlab}[1]{#1}
\providecommand{\url}[1]{\texttt{#1}}
\expandafter\ifx\csname urlstyle\endcsname\relax
  \providecommand{\doi}[1]{doi: #1}\else
  \providecommand{\doi}{doi: \begingroup \urlstyle{rm}\Url}\fi

\bibitem[Atan et~al.(2018)Atan, Jordon, and van~der Schaar]{atan2018deep}
O.~Atan, J.~Jordon, and M.~van~der Schaar.
\newblock Deep-treat: Learning optimal personalized treatments from
  observational data using neural networks.
\newblock In \emph{AAAI}, 2018.

\bibitem[Athey et~al.(2018)Athey, Imbens, and Wager]{athey2018approximate}
S.~Athey, G.~W. Imbens, and S.~Wager.
\newblock Approximate residual balancing: debiased inference of average
  treatment effects in high dimensions.
\newblock \emph{{Journal of the Royal Statistical Society: Series B
  (Statistical Methodology)}}, 80\penalty0 (4):\penalty0 597--623, 2018.

\bibitem[Belloni et~al.(2014)Belloni, Chernozhukov, and
  Hansen]{belloni2014inference}
A.~Belloni, V.~Chernozhukov, and C.~Hansen.
\newblock Inference on treatment effects after selection among high-dimensional
  controls.
\newblock \emph{The Review of Economic Studies}, 81\penalty0 (2):\penalty0
  608--650, 2014.

\bibitem[Bottou et~al.(2013)Bottou, Peters, Qui{\~n}onero-Candela, Charles,
  Chickering, Portugaly, Ray, Simard, and Snelson]{bottou2013counterfactual}
L.~Bottou, J.~Peters, J.~Qui{\~n}onero-Candela, D.~X. Charles, D.~M.
  Chickering, E.~Portugaly, D.~Ray, P.~Simard, and E.~Snelson.
\newblock {Counterfactual reasoning and learning systems: The example of
  computational advertising}.
\newblock \emph{JMLR}, 14\penalty0 (1):\penalty0 3207--3260, 2013.

\bibitem[B{\"u}hlmann(2013)]{buhlmann2013causal}
P.~B{\"u}hlmann.
\newblock Causal statistical inference in high dimensions.
\newblock \emph{{Mathematical Methods of Operations Research}}, 77\penalty0
  (3):\penalty0 357--370, 2013.

\bibitem[Dalessandro et~al.(2012)Dalessandro, Perlich, Stitelman, and
  Provost]{dalessandro2012causally}
B.~Dalessandro, C.~Perlich, O.~Stitelman, and F.~Provost.
\newblock {Causally motivated attribution for online advertising}.
\newblock In \emph{Workshop on Data Mining for Online Advertising and Internet
  Economy}, pages 1--9, 2012.

\bibitem[Diemert et~al.(2017)Diemert, Meynet, Galland, and
  Lefortier]{DiemertMeynet2017}
E.~Diemert, J.~Meynet, P.~Galland, and D.~Lefortier.
\newblock {Attribution modeling increases efficiency of bidding in display
  advertising}.
\newblock In \emph{Proceedings of the ADKDD'17}, pages 1--6, 2017.

\bibitem[Fan et~al.(2016)Fan, Imai, Liu, Ning, and Yang]{fan2016improving}
J.~Fan, K.~Imai, H.~Liu, Y.~Ning, and X.~Yang.
\newblock Improving covariate balancing propensity score: A doubly robust and
  efficient approach.
\newblock Technical report, Technical report, Princeton Univ, 2016.

\bibitem[Fong et~al.(2018)Fong, Hazlett, Imai, et~al.]{fong2018covariate}
C.~Fong, C.~Hazlett, K.~Imai, et~al.
\newblock {Covariate balancing propensity score for a continuous treatment:
  Application to the efficacy of political advertisements}.
\newblock \emph{The Annals of Applied Statistics}, 12\penalty0 (1):\penalty0
  156--177, 2018.

\bibitem[Galagate(2016)]{galagate2016causal}
D.~Galagate.
\newblock \emph{{Causal Inference with a continuous treatment and outcome:
  Alternative Estimators for parametric dos-response functions with
  Applications.}}
\newblock PhD thesis, 2016.

\bibitem[Guo et~al.(2016)Guo, Xue, and Chen]{guo2016cbps}
D.~Guo, L.~Xue, and H.~Chen.
\newblock Cbps-based inference in nonlinear regression models with missing
  data.
\newblock \emph{Open Journal of Statistics}, 6\penalty0 (4):\penalty0 675--684,
  2016.

\bibitem[Hirano and Imbens(2004)]{hirano2004propensity}
K.~Hirano and G.~W. Imbens.
\newblock The propensity score with continuous treatments.
\newblock \emph{Applied Bayesian modeling and causal inference from
  incomplete-data perspectives}, 226164:\penalty0 73--84, 2004.

\bibitem[Ho et~al.(2007)Ho, Imai, King, and Stuart]{ho2007matching}
D.~E. Ho, K.~Imai, G.~King, and E.~A. Stuart.
\newblock Matching as nonparametric preprocessing for reducing model dependence
  in parametric causal inference.
\newblock \emph{Political analysis}, 15\penalty0 (3):\penalty0 199--236, 2007.

\bibitem[Imai and Ratkovic(2014)]{imai2014covariate}
K.~Imai and M.~Ratkovic.
\newblock Covariate balancing propensity score.
\newblock \emph{{Journal of the Royal Statistical Society: Series B
  (Statistical Methodology)}}, 76, 2014.

\bibitem[Johansson et~al.(2016)Johansson, Shalit, and
  Sontag]{johansson2016learning}
F.~Johansson, U.~Shalit, and D.~Sontag.
\newblock Learning representations for counterfactual inference.
\newblock In \emph{ICML}, pages 3020--3029, 2016.

\bibitem[Kennedy et~al.(2017)Kennedy, Ma, McHugh, and Small]{kennedy2017non}
E.~H. Kennedy, Z.~Ma, M.~D. McHugh, and D.~S. Small.
\newblock Non-parametric methods for doubly robust estimation of continuous
  treatment effects.
\newblock \emph{Journal of the Royal Statistical Society: Series B (Statistical
  Methodology)}, 79\penalty0 (4):\penalty0 1229--1245, 2017.

\bibitem[Newman(2012)]{Newman-2012}
D.~Newman.
\newblock {UCI Machine Learning Repository}, 2012.
\newblock URL \url{https://archive.ics.uci.edu/ml/datasets/Bag+of+Words}.

\bibitem[Ramachandra(2018)]{ramachandra2018deep}
V.~Ramachandra.
\newblock {Deep Learning for Causal Inference}.
\newblock \emph{arXiv preprint arXiv:1803.00149}, 2018.

\bibitem[Robins et~al.(1994)Robins, Rotnitzky, and Zhao]{robins1994estimation}
J.~M. Robins, A.~Rotnitzky, and L.~P. Zhao.
\newblock Estimation of regression coefficients when some regressors are not
  always observed.
\newblock \emph{Journal of the American statistical Association}, 89\penalty0
  (427):\penalty0 846--866, 1994.

\bibitem[Schwab et~al.(2018)Schwab, Linhardt, and Karlen]{schwab2018perfect}
P.~Schwab, L.~Linhardt, and W.~Karlen.
\newblock {Perfect Match: A simple method for learning representations for
  counterfactual inference with neural networks}.
\newblock \emph{arXiv preprint arXiv:1810.00656}, 2018.

\bibitem[Schwab et~al.(2019)Schwab, Linhardt, Bauer, Buhmann, and
  Karlen]{schwab2019learning}
P.~Schwab, L.~Linhardt, S.~Bauer, J.~M. Buhmann, and W.~Karlen.
\newblock Learning counterfactual representations for estimating individual
  dose-response curves.
\newblock \emph{arXiv preprint arXiv:1902.00981}, 2019.

\bibitem[Shalit et~al.(2017)Shalit, Johansson, and Sontag]{pmlr-v70-shalit17a}
U.~Shalit, F.~D. Johansson, and D.~Sontag.
\newblock Estimating individual treatment effect: generalization bounds and
  algorithms.
\newblock In \emph{ICML}, volume~70, pages 3076--3085. PMLR, 2017.

\bibitem[Sharma et~al.(2019)Sharma, Gupta, Prasad, Chatterjee, Vig, and
  Shroff]{sharma2019metaci}
A.~Sharma, G.~Gupta, R.~Prasad, A.~Chatterjee, L.~Vig, and G.~Shroff.
\newblock {MetaCI: Meta-Learning for Causal Inference in a Heterogeneous
  Population}.
\newblock \emph{{NeurIPS CausalML workshop; arXiv preprint arXiv:1912.03960}},
  2019.

\bibitem[Sharma et~al.(2020)Sharma, Gupta, Prasad, Chatterjee, Vig, and
  Shroff]{esann-multiMBNN}
A.~Sharma, G.~Gupta, R.~Prasad, A.~Chatterjee, L.~Vig, and G.~Shroff.
\newblock {MultiMBNN: Matched and Balanced Causal Inference with Neural
  Networks}.
\newblock \emph{{ESANN; arXiv preprint arXiv:2004.13446}}, 2020.

\bibitem[Sun et~al.(2015)Sun, Wang, Yin, Yang, and Chang]{sun2015causal}
W.~Sun, P.~Wang, D.~Yin, J.~Yang, and Y.~Chang.
\newblock Causal inference via sparse additive models with application to
  online advertising.
\newblock In \emph{AAAI}, 2015.

\bibitem[Weinstein et~al.(2013)Weinstein, Collisson, Mills, Shaw, Ozenberger,
  Ellrott, Shmulevich, Sander, Stuart, Network, et~al.]{weinstein2013cancer}
J.~N. Weinstein, E.~A. Collisson, G.~B. Mills, K.~R.~M. Shaw, B.~A. Ozenberger,
  K.~Ellrott, I.~Shmulevich, C.~Sander, J.~M. Stuart, C.~G. A.~R. Network,
  et~al.
\newblock The cancer genome atlas pan-cancer analysis project.
\newblock \emph{{Nature Genetics}}, 45\penalty0 (10):\penalty0 1113, 2013.

\bibitem[Zhang et~al.(2019)Zhang, Lan, Ding, Wang, Hassanpour, and
  Greiner]{zhang2019reducing}
Z.~Zhang, Q.~Lan, L.~Ding, Y.~Wang, N.~Hassanpour, and R.~Greiner.
\newblock Reducing selection bias in counterfactual reasoning for individual
  treatment effects estimation.
\newblock \emph{arXiv preprint arXiv:1912.09040}, 2019.

\end{thebibliography}

\end{document}